\documentclass{taira}
\usepackage{graphicx}
\usepackage{epstopdf, epsfig}
\usepackage{graphicx}
\usepackage{dcolumn}
\usepackage{bm}
\usepackage{color}
\usepackage{array}
\usepackage{tabu}
\usepackage{amsmath,amssymb,stmaryrd}
\usepackage{multicol}
\usepackage{algorithm}
\usepackage[noend]{algpseudocode}
\usepackage{subfigure}
\usepackage{subfigmat,comment}
\usepackage{pbox}
\usepackage{xfrac}
\usepackage{tabularx}
\usepackage{multirow}
\usepackage[percent]{overpic}
\usepackage{float}
\usepackage{xcolor}
\usepackage{tikz}
\usetikzlibrary{shapes}
\usepackage{tikz}

\begin{document}  

\newcommand{\sqdiamond}[1][fill=black]{\tikz [x=1.2ex,y=1.85ex,line width=.1ex,line join=round, yshift=-0.285ex] \draw  [#1]  (0,.5) -- (.5,1) -- (1,.5) -- (.5,0) -- (0,.5) -- cycle;}%
\newcommand{\MyDiamond}[1][fill=black]{\mathop{\raisebox{-0.275ex}{$\sqdiamond[#1]$}}}

\newcommand{\markerone}{\raisebox{0.0pt}{\tikz{\node[scale=0.6,regular polygon, circle, draw = {rgb,255:red,0; green,114; blue,189},line width=0.3mm, fill={rgb,255:red,255; green,255; blue,255}](){};}}}
\newcommand{\markertwo}{\raisebox{0.0pt}{\tikz{\node[scale=0.45, regular polygon, regular polygon sides=3, fill={rgb,255:red,126; green,47; blue,142}](){};}}}
\newcommand{\markerthree}{\raisebox{0.0pt}{\tikz{\node[scale=0.45,regular polygon, regular polygon sides=3,fill={rgb,255:red,119; green,172; blue,48},rotate=-90](){};}}}
\newcommand{\markerfour}{\raisebox{0.0pt}{\tikz{\node[scale=0.45,regular polygon, regular polygon sides=3,fill={rgb,255:red,77; green,190; blue,238},rotate=-180](){};}}}
\newcommand{\markerfive}{\raisebox{0.0pt}{\tikz{\node[scale=0.55,regular polygon, regular polygon sides=4, draw = {rgb,255:red,237; green,177; blue,32}, line width=0.3mm, fill={rgb,255:red,255; green,255; blue,255},rotate=0](){};}}}
\newcommand{\markersix}{\raisebox{0.0pt}{\tikz{\node[scale=0.55,regular polygon, regular polygon sides=4, draw={rgb,255:red,217; green,83; blue,25}, fill={rgb,255:red,255; green,255; blue,255},rotate=-45](){};}}}

\newtheorem{lemma}{Lemma}
\newtheorem{corollary}{Corollary}
\newcommand{\bs}{\boldsymbol}
\shorttitle{Laminar wakes of swept wings} 
\shortauthor{K. Zhang et al.} 

\title{Laminar separated flows over finite-aspect-ratio swept wings}
\author
 {
 Kai Zhang\aff{1}
  \corresp{Present address for correspondence: Department of Mechanical and Aerospace Engineering, Rutgers University. Email: kai.zhang3@rutgers.edu},
  Shelby Hayostek\aff{2},
  Michael Amitay\aff{2},
  Anton Burtsev\aff{3},
  Vassilios Theofilis\aff{3,4}, 
  \and
  Kunihiko Taira\aff{1}
  }

\affiliation
{
\aff{1}
Department of Mechanical and Aerospace Engineering, University of California, Los Angeles, CA 90095, USA

\aff{2}
Department of Mechanical, Aeronautical, and Nuclear Engineering, Rensselaer Polytechnic Institute, Troy, NY 12180, USA

\aff{3}
Department of Mechanical, Materials and Aerospace Engineering, University of Liverpool, Brownlow Hill, England L69 3GH, United Kingdom

\aff{4}
Department of Mechanical Engineering, Escola Politecnica, Universidade S\~ao Paulo, Avda. Prof. Mello Moraes 2231,  CEP 5508-900, S\~ao Paulo, Brasil 
}
\maketitle
\begin{abstract}
We perform direct numerical simulations of laminar separated flows over finite-aspect-ratio swept wings at a chord-base Reynolds number of $Re = 400$ to reveal a variety of wake structures generated for a range of aspect ratios ($sAR=0.5-4$), angles of attack ($\alpha=16^{\circ}-30^{\circ}$), and sweep angles ($\Lambda=0^{\circ}-45^{\circ}$).  
Flows behind swept wings exhibit increased complexity in their dynamical features compared to unswept-wing wakes.
For unswept wings, the wake dynamics are predominantly influenced by the tip effects. Steady wakes are mainly limited to low-aspect-ratio wings. Unsteady vortex shedding takes place near the midspan of higher-$AR$ wings due to weakened downwash induced by the tip vortices.  
With increasing sweep angle, the source of three dimensionality transitions from the tip to the midspan.
A pair of symmetric vortical structures forms along the two sides of the midspan, imposing downward velocity upon each other in a manner similar to the downwash induced by tip vortices.
Such stabilizing midspan effects not only expand the steady wake region to higher aspect ratios, but also enhance lift.
At higher aspect ratios, the midspan effects of swept wings diminish at outboard region, allowing unsteady vortex shedding to develop near the tip.
In the wakes of highly swept wings, streamwise finger-like structures form repetitively along the wing span, providing a stabilizing effect.
The insights revealed from this study can aid the design of high-lift devices and serve as a stepping stone for understanding the complex wake dynamics at higher Reynolds numbers and those generated by unsteady wing maneuvers.

\end{abstract}

\section{Introduction}
\label{sec:intro}

Separated flows over lifting surfaces have been studied extensively due to their critical importance in aerodynamics and hydrodynamics.
Over the past few decades, substantial studies have been dedicated to the understanding of post-stall flows over finite-aspect-ratio wings \citep{winkelmann1980effects,taira2009three,mulleners2017flow,eldredge2019leading,zhang2020formation}.
For a steadily translating wing, the vortices generated from the leading and trailing edges exhibit complex nonlinear evolution under the influence of the tip vortices. The resulting wake is highly three-dimensional in nature.

The introduction of sweep to the finite-aspect-ratio wings further enriches the wake dynamics. 
\citet{harper1964review} asserted that once local stall appears on the swept wing, the spanwise boundary layer flow alters the stall characteristics of the attached sections of the span. Thus the separated flows over swept wings bear little resemblance to the two-dimensional flows.
The stalled flow over a swept wing usually features a ``ram's horn" vortex, which stems from the inboard leading edge and grows in size as it trails to the wake behind the tip. Such unique flow is also referred to as ``tip stall" \citep{black1956flow,zhang2019experimental}.
Focusing on the detailed flow structures,
\citet{yen2007flow} and \citet{yen2009flow} experimentally studied the effects of sweep angle, Reynolds number and angle of attack on the wake vortices, boundary layer flow patterns and aerodynamic performance of the swept wings. The wake features were found to be significantly dependent on these parameters.
\citet{visbal2019effect} conducted large eddy simulations of flows over swept wings with a semi aspect ratio of 4 at $Re=2\times 10^{5}$. A region of laminar flow is observed near the midspan, and it grows in extent with increasing sweep angle. The authors attributed this observation to the relief effect provided by the sweep-induced spanwise flow toward the tip. 
Furthermore, dynamic stall was numerically simulated for a swept wing undergoing pitching and plunging maneuvers \citep{visbal2019effect,garmann2020examination}. Under these wing motions, the release of the arch-shaped vortices during downstroke occurs near both tip regions, whereas only a single arch vortex is seen to detach from the midspan in the unswept case. 

As a special case, delta wing is well known for harnessing separated flow physics to enhance its aeordynamic performance. 
Unlike the unswept wing, for which the leading-edge vortex grows and inevitably sheds away, the LEVs on the delta wing are stable due to the balance between spanwise vorticity transport and local vorticity generation \citep{polhamus1966concept,rockwell1993three,gursul2005unsteady}.
As the LEVs trail downstream, they create low pressure on the suction side of the wing, producing a sizable vortical lift.
Such favorable effect grants delta wing the ability to operate at much higher angle of attack than a conventional wing.

The LEVs may also contribute to enhanced lift for sustained avian flight. 
\citet{videler2004leading} identified LEVs in the swept wings of common swifts (\emph{Apus apus}) during gliding.
\citet{ben2019lift} proposed a pseudo-three-dimensional flow model for investigating the stationary LEV mechanism over swept back wings. The model revealed that wing geometry has a major role in the localization of the stationary LEVs over high aspect-ratio wings.
In addition to the above studies, LEVs on translating swept wings has been further examined by \citet{lentink2009rotational} and \citet{beem2012stabilization}, who showed that the spanwise flow alone can not sustain stationary LEVs.
Moreover, the effectiveness of the LEVs in augmenting lift is still arguable for the low-speed flights \citep{lentink2007swifts}. 
These mixed findings regarding the role of separated flows in avian flights with swept wings warrants further investigations.


The detailed features of separated flows over swept wings depends on various parameters including the aspect ratio, angle of attack, and sweep angle.
The interplay between these effects generates complex wake dynamics, which are not thoroughly understood thus far.
In this work, we present an extensive numerical study on the laminar separated flows over swept wings. 
The objective is to characterize the different wake structures observed over a large range of parameters, and to identify their formation mechanisms.
The rest of the paper is organized as follows. In what follows, we present the computational setup in \S \ref{sec:setup}. A variety of wakes observed from this study is reported in \S \ref{sec:results}. We conclude this study by summarizing our findings in \S \ref{sec:conclusion}.

\section{Computational setup}
\label{sec:setup}

\begin{figure}
\centering
\includegraphics[scale=0.45]{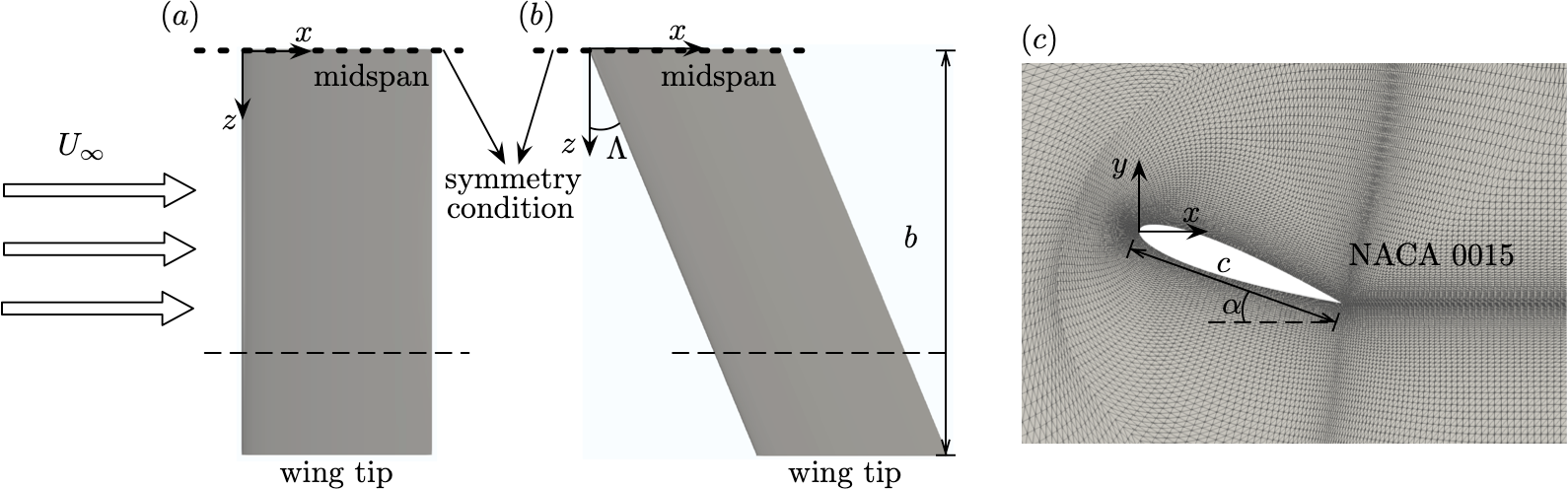}
\caption{Schematic of setup for $(a)$ unswept wing and $(b)$ swept wing. $(c)$ shows the cross-sectional slice along the broken lines in $(a)$ and $(b)$.}
\label{fig:scheme}
\end{figure}

We consider three-dimensional incompressible flows over swept finite-aspect-ratio wings with the NACA 0015 cross-section profile. 
The setup of the wing geometry is shown in figure \ref{fig:scheme}. 
The wings are subjected to uniform flow with velocity $U_{\infty}$ in the $x$ direction. The $z$ axis aligns with the spanwise direction of an unswept wing, and the $y$ axis points at the lift direction. 
For the swept wings, the sweep angle $\Lambda$ is defined as the angle between the $z$ axis and leading edge of the wing. 
We consider a range of sweep angles from $0^{\circ}$ to $45^{\circ}$, at an interval of $7.5^{\circ}$.
Half of the swept wing model is simulated by prescribing a symmetry boundary condition along the midspan. 
Denoting the half wing span as $b$ and the chord length as $c$, the semi aspect ratio is defined as $sAR=b/c$, which is varied from 0.5 to 4. 
We focus on flows that develop behind wings at high angles of attack, $\alpha=16^{\circ}$, $20^{\circ}$, $26^{\circ}$ and $30^{\circ}$, as we are particularly interested in separated flows.
The Reynolds number, which is defined as $Re= U_{\infty}c/\nu$ ($\nu$ is the kinematic viscosity), is kept fixed at 400.
In what follows, all spatial variables are scaled by the chord length $c$, velocity by freestream velocity $U_{\infty}$, and time by $c/U_{\infty}$. 

The flows over swept wings are simulated by numerically solving the three-dimensional Navier-Stokes equations. An incompressible solver \emph{Cliff} (in \emph{CharLES} software package, Cascade Technologies, Inc.) is used for the direct numerical simulations. The solver employs a collocated, node-based finite-volume method to compute the solutions to the governing equations with second-order accuracy in both space and time \citep{ham2004energy,ham2006accurate}. The computational domain and the spatial discretization setup in this study follow our previous work with extensive validation \citep{zhang2020formation}. For the swept cases, the straight wing mesh system is sheared in the $x$ direction along with the wing. 

\section{Results}
\label{sec:results}
\begin{figure}
	\centering
	\includegraphics[width=0.95\textwidth]{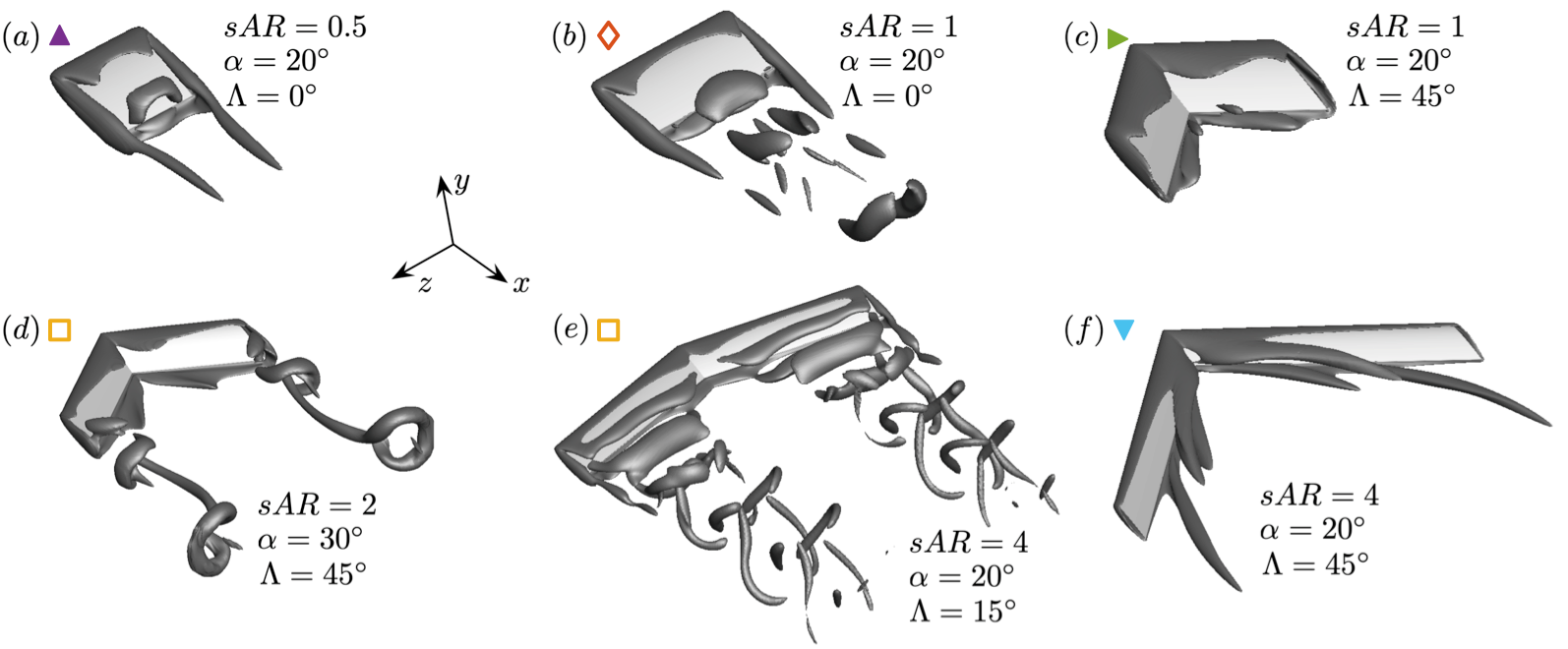}
	\caption{Representative wakes of swept wings. $(a)$ Steady flow with tip vortex \protect\markertwo;  $(b)$ unsteady shedding near midspan $\MyDiamond[draw={rgb,255:red,217; green,83; blue,25},line width=0.3mm, fill=white]$;  $(c)$ steady flow with midspan structures \protect\markerthree
; $(d)$ and $(e)$ unsteady shedding near wing tip \protect\markerfive; $(f)$ steady flow with streamwise vortices \protect\markerfour. The figures are scaled for visual clarity.  
}
	\label{fig:VorticalStructures}
\end{figure}

\begin{figure}
	\centering
	\includegraphics[width=0.99\textwidth]{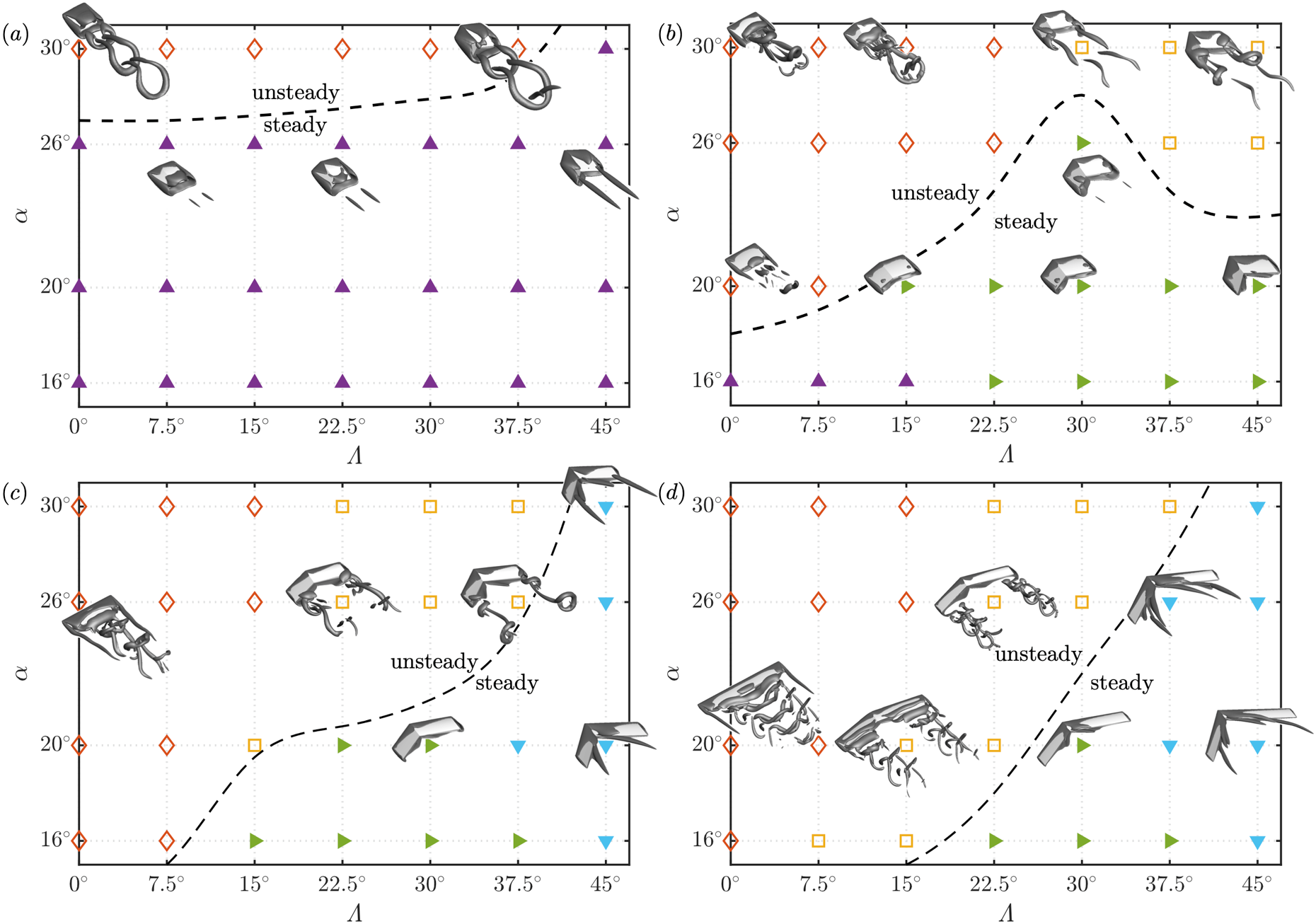}
	\caption{Classification of vortical structures behind finite-aspect-ratio swept wings. $(a)$ $sAR=0.5$; $(b)$ $sAR=1$; $(c)$ $sAR=2$; and $(d)$ $sAR=4$. The dash lines denote the approximate boundaries between steady (filled symbols) and unsteady (empty symbols) flows over the $\Lambda$-$\alpha$ space. Vortical structures are visualized by iso-surfaces of $Q=1$ for select cases.}
	\label{fig:regimes}
\end{figure}

The wakes of swept wings exhibit a rich variety of  features depending on the aspect ratio, angle of attack, and sweep angle.
We show representative wakes in figure \ref{fig:VorticalStructures}, with their distributions over the $\Lambda$-$\alpha$ space in figure \ref{fig:regimes} for different aspect ratios. 
The wakes are broadly divided into two categories: steady flows and unsteady flows.
Steady flows take different forms, including those with tip vortices (\protect\markertwo), those with midspan structures (\protect\markerthree), and those with streamwise vortices (\protect\markerfour). The steady flow regime appear over a wide range of the parameter space.
Unsteady flows are characterized by vortex shedding near the midspan ($\MyDiamond[draw={rgb,255:red,217; green,83; blue,25},line width=0.3mm, fill=white]$) or near the tip (\protect\markerfive).
In what follows, we describe key mechanisms that are responsible for the formation of the different wakes mentioned above.

\subsection{Tip effects}
\label{subsec:tipvortex}

Tip effects play an important role in the development of wakes behind low-aspect-ratio wings. 
At $sAR=0.5$, steady flows are observed up to $\alpha=26^{\circ}$, as shown in figure \ref{fig:regimes}$(a)$. 
These steady flows typically feature a pair of tip vortices, which induce strong downwash over the wing span, suppressing the formation of large leading-edge vortices.
Such mechanism is responsible for the stability of the wake, particularly for low-aspect-ratio unswept wings \citep{taira2009three,devoria2017mechanism,zhang2020formation}.
At higher sweep angles, the vortical structures emanated from the leading edge grows in size, covering a significant portion of the suction surface of the wing. 
These vortical structures are also beneficial to the stability of the wake, as will be discussed in detail in \S \ref{subsec:midspan}. 
Unsteady flows are only observed at  $\alpha\gtrsim 30^{\circ}$ for $\Lambda=0^{\circ}-37.5^{\circ}$. 
These unsteady wakes are characterized by the periodic shedding of hairpin vortices \citep{taira2009three}. 

As the aspect ratio is increased, the downwash induced by the tip vortices weakens along the midspan, allowing the roll-up of vortex sheet at the leading edge.
As a result, the unsteady vortex shedding develops near the midspan, as visualized in figure \ref{fig:VorticalStructures}$(b)$.
The stability boundary (dash line in figure \ref{fig:regimes}) shifts towards lower angles of attack at $sAR=1$ and 2 for low-swept wings. 
Additional details on the three-dimensional unsteady wake dynamics of wings under tip effects are reported in \citet{taira2009three} and \citet{zhang2020formation}.

\subsection{Midspan effects}
\label{subsec:midspan}

\begin{figure}
\centering
\includegraphics[scale=0.47]{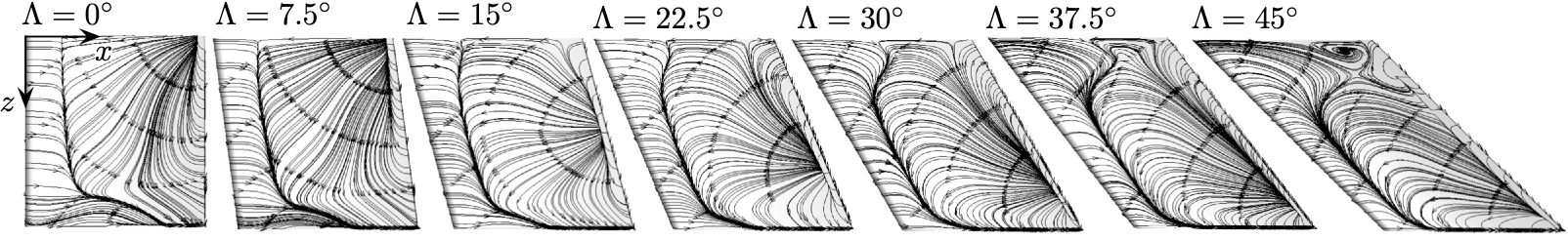}
\caption{Skin-friction lines on the suction side of wings with $(\alpha,sAR)=(20^{\circ},1)$}
\label{fig:SkinFrictionLine}
\end{figure}

For wings with larger aspect ratio ($sAR\gtrsim 1$), in place of the weakened tip effects, the midspan symmetry introduces another type of three dimensionality that dominates the wake. 
Let us present the skin-friction lines for cases of $(\alpha,sAR)=(20^{\circ},1)$ with varying $\Lambda$ as shown in figure \ref{fig:SkinFrictionLine}. 
With the increase in sweep angle, the boundary layer separation point near the midspan gradually shifts towards the trailing edge.
This is caused by the growth of the vortical structures emanated from the leading edge, as observed in figure \ref{fig:regimes}$(b)$.
The increase in sweep angle for $sAR\gtrsim 1$ also leads to the attenuation of the tip vortices, as reflected by the diminishing three-dimensional skin-friction line pattern near the wing tip in figure \ref{fig:SkinFrictionLine}. 
The above observations suggest a switch-over of the source of three-dimensionality from the wing tip in low-$\Lambda$ cases to the midspan in high-$\Lambda$ cases.

\begin{figure}
\centering
\includegraphics[width=0.97\textwidth]{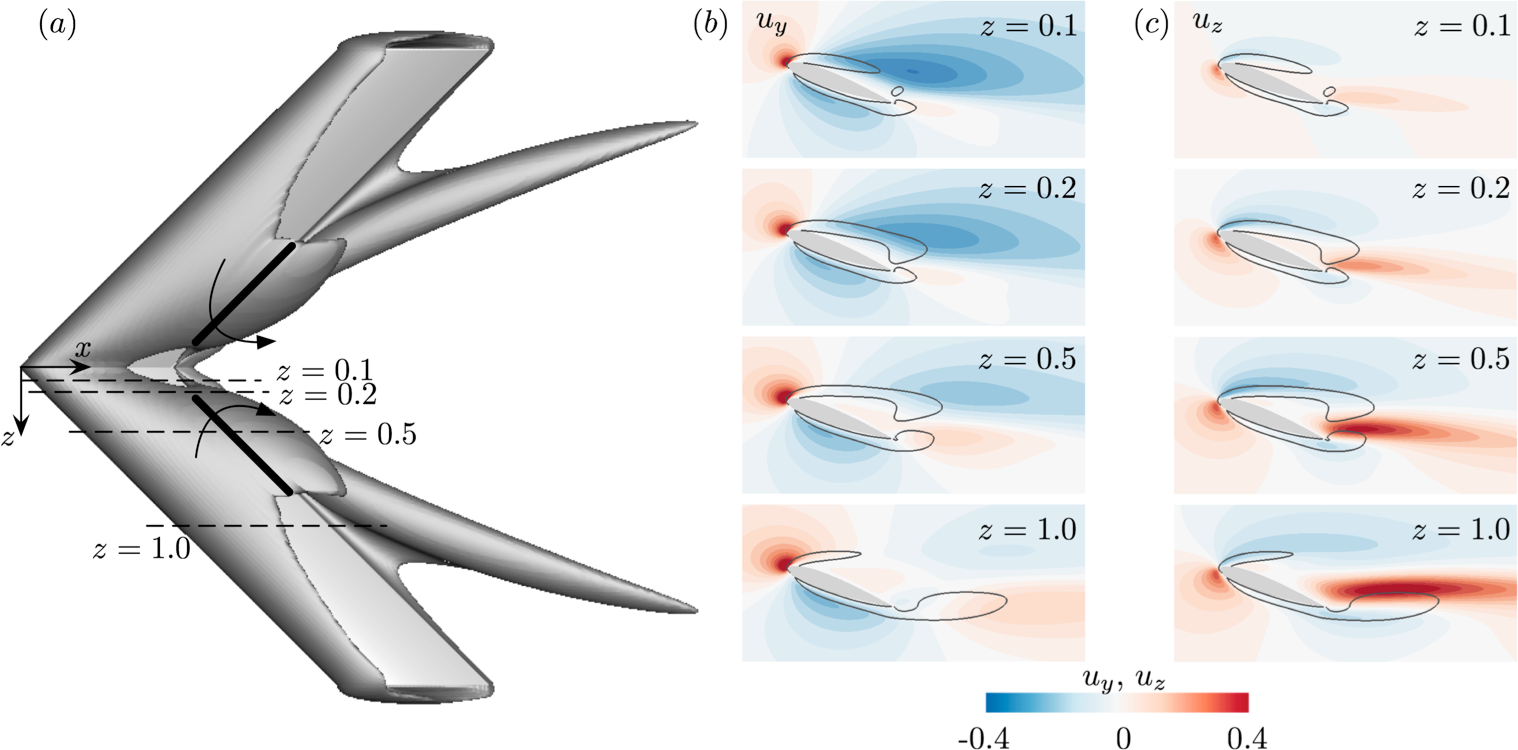}
\caption{Wake for the case $(sAR, \alpha, \Lambda)=(2, 20^{\circ}, 45^{\circ})$. $(a)$ Vortical structures visualized by isosurface of $Q=1$. The thick solid lines represent the approximate location of vortical structures near the midspan, with the curved arrows showing their directions of rotation. The dash lines indicates the locations of visualizations shown on the right. $(b)$: $u_y$ and $(c)$: $u_z$ fields at different $z$ locations.}
\label{fig:MidSpanDownwash}
\end{figure}

Let us take a closer look at the vortical structures near the midspan. A representative case of $(\alpha,sAR,\Lambda)=(20^{\circ}, 2, 45^{\circ})$ is shown in figure \ref{fig:MidSpanDownwash}. 
For this case, the vortex sheet rolls up along the spanwise direction, covering the entire chord over the inboard section of the wing.
Due to the symmetric condition, the identical vortical structures on two sides of the mid-plane are oriented at an angle of $180^{\circ} - 2\Lambda$, which in the current case is $90^{\circ}$. 
As a result, each of the vortical structures is subjected to the downward velocity (pointing at $-y$ direction) induced by its symmetric peer on the other side of the midspan.
This is clearly manifested in figure \ref{fig:MidSpanDownwash}$(b)$ by the strong negative crossflow velocity $u_y$ on the suction side near the $z=0$ plane.

Due to such three-dimensional midspan effects, the vortex sheet emanated from the leading edge is pinned to the suction side of the wing, forming steady vortical structures.
For cases with smaller $\Lambda$, the angle between the symmetric vortical structure tends towards $180^{\circ}$, and the mutual induced velocity becomes smaller. 
The three dimensionality developed from the midspan is able to stabilize the wakes of a considerable number of cases, as labeled by the green triangle (\protect\markerthree) in figure \ref{fig:regimes}.
The midspan effects are also observed for swept wings with $sAR=0.5$, in which cases the tip vortices also stabilize the wake. 
In fact, the strengthening of the tip vortices with increasing $\Lambda$ at $sAR=0.5$ (see figure \ref{fig:regimes}$(a)$, $\alpha=26^{\circ}$ for example) is likely related to the three-dimensional midspan effects. As vortex sheets emanated from leading edge are less likely to roll up due to the mutual downwash, they counteract less on the formation of tip vortices. 
We note that spanwise vorticity transport, which is considered as the key mechanism for the formation of steady LEVs on delta wings \citep{polhamus1966concept,gursul2005review}, does not play an important role in the formation of the midspan vortical structures in the current case. This is manifested by the fact that the strong outboard velocity $u_z$ does not coincide with the vortex core, as shown in figure \ref{fig:MidSpanDownwash}$(c)$.


The downward induced velocity described above is strong near the midspan, as shown in figure \ref{fig:MidSpanDownwash}$(b)$.
With the gradual weakening of the midspan effects towards the outboard sections, unsteadiness develops locally near the tip region, while the midspan region still remains steady. The resulting flows resemble the ``tip stall" phenomenon, as described in \citet{black1956flow,zhang2019experimental,visbal2019effect}. 
This type of flows, as indicated by yellow squares (\protect\markerfive) in figure \ref{fig:regimes}, prevails for swept wings with large aspect ratios (e.g., $sAR=4$) and high angles of attack (e.g., $\alpha=26^{\circ}-30^{\circ}$).
For wings with large $sAR$, the unsteady shedding develops from the leading edge and features mostly spanwise vortices, see figure \ref{fig:VorticalStructures}$(e)$ for example.
In contrast, for wings with low aspect ratio but large angles of attack, the unsteady vortices are shed from the wing tip, and they feature hairpin structures with the legs originated from the pressure and suction sides of the wing, as shown in figure \ref{fig:VorticalStructures}$(f)$.

\subsection{Formation of streamwise finger-like structures}
\label{subsec:streamwise}

\begin{figure}
\centering
\includegraphics[width=0.97\textwidth]{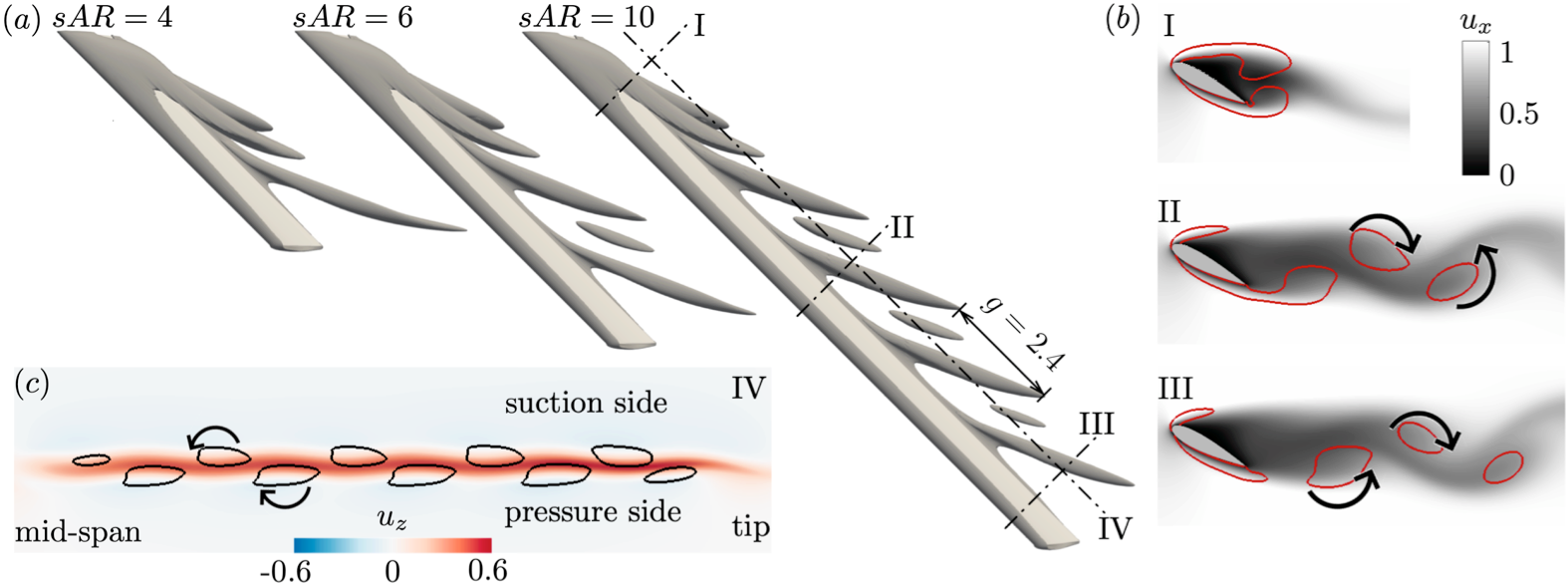}
\caption{Streamwise vortices along high-aspect-ratio wings with $\Lambda=45^{\circ}$. $(a)$ Vortical structures visualized by isosurface of $Q=1$ for $sAR=4$, 6 and 10. $(b)$ $u_x$ fields at indicated slices as shown in $(a)$. The red lines represent contours of $Q=0.5$ with the curved arrows indicating the direction of rotation. $(c)$ $u_z$ field on the slice IV. Black circles show the isosurfaces of $Q=0.5$.}
\label{fig:StreamwiseVortices}
\end{figure}

For high sweep angles of $\Lambda \gtrsim 37.5^{\circ}$, the flows over high-aspect-ratio wings ($sAR \gtrsim 2$) transition from the unsteady tip shedding to steady wakes, through the formation of the streamwise finger-like structures.
As shown in figure \ref{fig:VorticalStructures}$(f)$, the finger-like structures are oriented at an angle higher than $\Lambda$ with respect to the $z$ axis.
These structures bend towards the streamwise direction away from the wing. 
Wakes of swept wings with even $sAR$ as large as 10 are stabilized with the alternating formation of the streamwise structures along the wing span, as shown in figure \ref{fig:StreamwiseVortices}$(a)$.

These streamwise vortices observed in the present work are formed under the same mechanism with those in the wakes of axisymmetric slender bodies at high incidence \citep{sarpkaya1966separated,thomson1971spacing}.
The impulse flow analogy has long been used to understand these vortical structures. 
According to this analogy, the progressive development of the wake along the wing span when viewed in cross-flow planes is similar to the temporal growth of the flow behind a two-dimensional wing translated impulsively from rest. 
Three slices at different spanwise locations in figure \ref{fig:StreamwiseVortices}($b$) show a temporal-like evolution of wake vortices along the wing span.
In addition, the spacing between the neighbouring streamwise vortices is fixed at $g = 2.4$, which follows $g \approx  U_{\infty}\sin\Lambda\cdot T_{2D}$, in which $U_{\infty}\sin\Lambda$ represents the speed of the spanwise flow, and $T_{2D}=3.2$ is the nondimensional period of vortex shedding in the analogous two-dimensional case (for an infinite swept wing). 
This relationship suggests a close analogy between the two-dimensional unsteady flow and the present three-dimensional steady flow, where the time-dependence of the former is replaced by spatial dependence of the latter.

From another perspective, when viewed from the wake on a plane parallel to the wing span, as shown in figure \ref{fig:StreamwiseVortices}$(c)$, the steady streamwise vortices are positioned at two sides of strong outboard velocity $u_z$, resembling the configuration where inverse K\'arm\'an vortices are formed in the jet profile.
This observation suggests the important role of the spanwise velocity in the formation of steady three-dimensional vortices. 
Similar streamwise structures have also been reported for rotating large-$AR$ wings \citep{jardin2017coriolis}, in which spanwise flow is promoted by artificially adding Coriolis force.


\subsection{Aerodynamic forces}
\label{subsec:forces}

\begin{figure}
\centering
\includegraphics[width=0.99\textwidth]{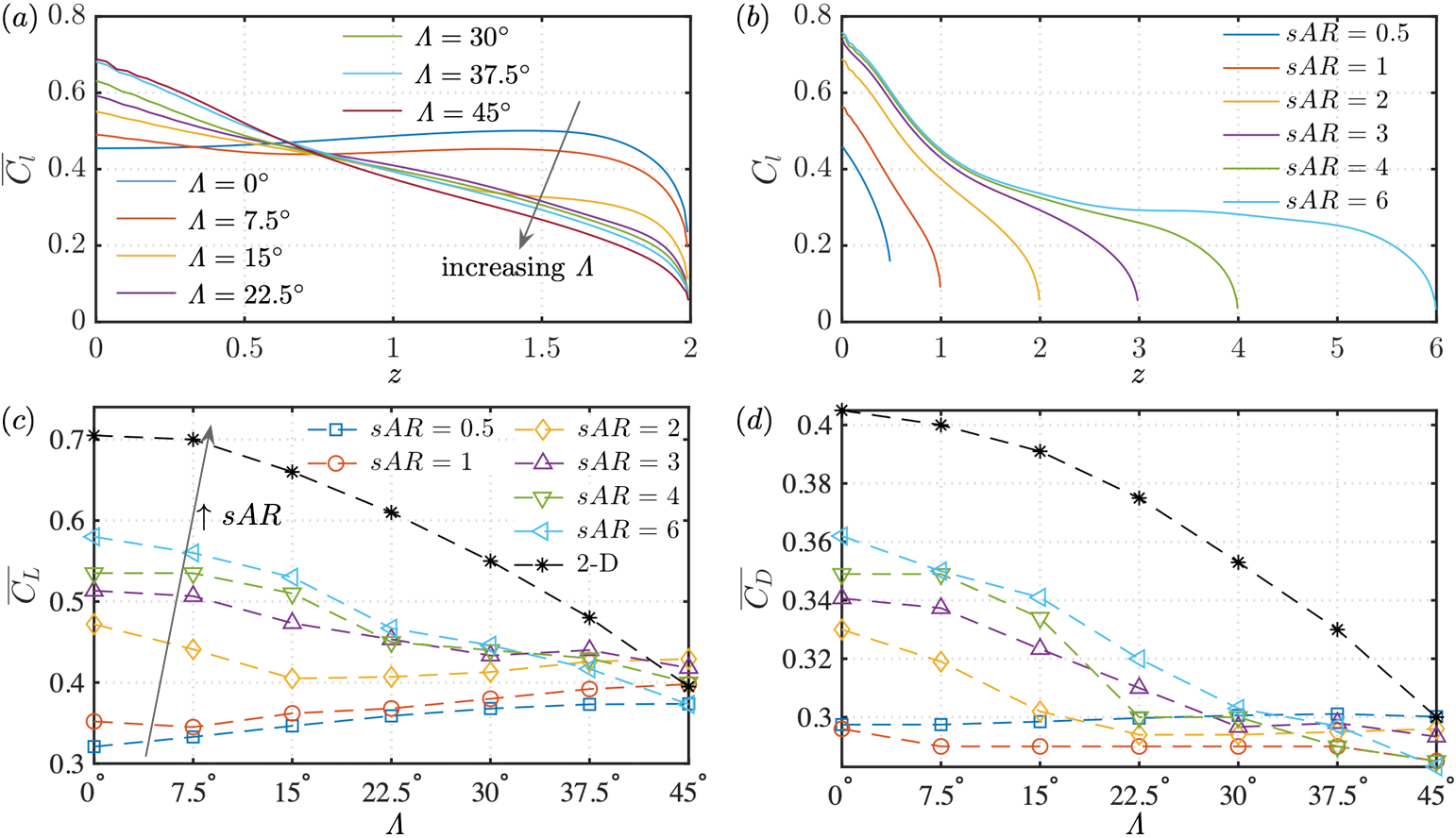}
\caption{Force coefficients for $\alpha=20^{\circ}$. $(a)$ Sectional lift coefficients for varied sweep angles at a fixed aspect ratio $sAR=2$. $(b)$ Sectional lift coefficients for cases of $\Lambda=45^{\circ}$ with different aspect ratios. $(c)$ and $(d)$ lift coefficients $\overline{C_L}$ and drag coefficients $\overline{C_D}$ over sweep angle $\Lambda$ for different aspect ratios.}
\label{fig:CLCD}
\end{figure}

The vortical structures described above influence the aerodynamic forces on the swept wings. We examine the effects of sweep angle on the distribution of the time-averaged sectional lift coefficients $\overline{C_l}$ for the representative cases of $(\alpha,sAR)=(20^{\circ},2)$ in figure \ref{fig:CLCD}$(a)$. 
For the unswept wing, the sectional lift increases slowly from the midspan towards outboard, reaching maximum at $z\approx 1.5$ before its drastic decrease at the tip. 
The slight swell-up of the sectional lift at the outboard location is attributed to downwash from the tip vortices \citep{devoria2017mechanism,zhang2020formation}.
For swept wings, the sectional lift coefficients are significantly higher at the inboard locations, due to the additional circulation maintained by the midspan effects. 
We note that the lift distribution of swept wings in high $Re$ flows exhibits the opposite pattern, where sectional lift is low near midspan and high at outboard sections \citep{schlichting1959aerodynamik,nickel1994tailless}. Moreover, for the same angle of attack, one sees lift loss everywhere along the span of swept wing compared with the unswept wing \citep{visbal2019effect}. 
These differences highlight the midspan effects in augmenting lift of swept wings in low Reynolds number separated flows.

The effects of aspect ratio on the distribution of sectional lift coefficients of swept wings are examined for the cases $(\alpha,\Lambda)=(20^{\circ},45^{\circ})$ in figure \ref{fig:CLCD}$(b)$. 
Monotonic decrease of sectional lift is observed across $sAR=0.5$ to 6. 
The maximum sectional lift at midspan $z=0$ increases with the aspect ratio, and eventually saturates as $sAR$ reaches 3.
This observation suggests that the tip effect is still noticeable
for low-$AR$ swept wings even though distinct tip vortices are not observed.
For high-$AR$ wings, the $C_l$-$z$ curves feature steep slope over the inboard region, reflecting the weakening of the midspan effects. The slope becomes much gentler towards the outboard, where midspan effects diminish. 

Finally, we present the total lift $\overline{C_L}$ and drag $\overline{C_D}$ coefficients of the swept wings at $\alpha=20^{\circ}$ with different aspect ratios and varying sweep angles in figure \ref{fig:CLCD}$(c)$ and $(d)$.
For swept wings with $sAR=0.5$ and 1, the lift coefficients increase with sweep angle, due to the dominance of the midspan effects over the entire wing span. 
The drag coefficients of the low-aspect-ratio wings, on the other hand, do not show significant variations over the increase of sweep angle.
It is interesting to note that $\overline{C_D}$ for $sAR=0.5$ is higher than that for $sAR=1$, an observation also reported in \citet{taira2009three}.
For wings with $sAR\gtrsim 3$, the contribution of the high sectional lift at inboard region to the total lift force is overshadowed by the lower sectional lift at the outboard region.
As a result, the lift coefficients of wings with high aspect ratio decrease with increasing sweep angle.
Similar trend is also observed for the drag coefficients of  high-aspect-ratio wings.
While for wings with small $\Lambda$ the lift coefficients $\overline{C_L}$ are positively related with $sAR$, at the highest sweep angle of $\Lambda=45^{\circ}$, maximum lift coefficient is achieved with $sAR=2$, and is even higher than the analogous two-dimensional case.
The midspan effects as lift enhancement mechanism for low-aspect-ratio swept wings could potentially inspire designs of high lift devices.

\section{Conclusion}
\label{sec:conclusion}

We have studied laminar separated flows over finite-aspect-ratio swept wings with direct numerical simulations at a chord-based Reynolds number of 400. 
Due to the complex interplay between the effects of aspect ratio, sweep angle and angle of attack, the wakes of finite swept wings exhibit a variety of vortical features, which are not observed behind the unswept wings.
We have described key mechanisms that are responsible for the emergence of different types of flows over swept wings. 
For wings with low aspect ratios and low sweep angles, the downwash by the tip vortices stabilizes the wake. 
With the increase in aspect ratio, the downwash weakens along the midspan, allowing the formation of unsteady vortex shedding. 
For higher sweep angles, the source of three dimensionality in the wake transitions from the wing tip to the midspan.
A pair of symmetric vortical structures form near the midspan due to their mutually induced downward velocity, which stabilizes the wake of higher-aspect-ratio swept wings.
Such midspan effects also act as a lift enhancement mechanism for wings with low to medium aspect ratios.
At high aspect ratio, the midspan effects diminishes near the outboard of the wing, and unsteady vortex shedding occurs near the wing tip region. 
For wings with high aspect ratios, steady wakes are again achieved at high sweep angles, where a transposition occurs from two-dimensional unsteady flow to three-dimensional steady flow.
The resulting steady wake features the repetitive formation of the streamwise finger-like structures along the span. 

This study provided a detailed look into the effects of sweep on the wake dynamics of finite-aspect-ratio wings. The insights obtained from this study, particularly  those regarding the midspan effects, could potentially be used for designing high-lift devices. In addition, the knowledge gained here also forms a stepping stone for further understanding the complex wake dynamics at higher Reynolds numbers and those generated by unsteady wing maneuvers.

\section*{Declaration of interest}
The authors report no conflict of interest.

\section*{Acknowledgement}
We acknowledge the generous support from the US Air Force Office of Scientific Research (FA9550-17-1-0222) monitored by Dr. Gregg Abate.

\bibliography{jfm-reference}
\bibliographystyle{jfm}
\end{document}